# Optical Response of Finite-Thickness Ultrathin Plasmonic Films

Igor V. Bondarev[1,*], Hamze Mousavi[1], and Vladimir M. Shalaev[2]

[1]*Math & Physics Department, North Carolina Central University, Durham, NC 27707, USA*

[2]*School of Electrical & Computer Engineering and Birck Nanotechnology Center, Purdue University, West Lafayette, IN 47907, USA*

**Abstract:** We discuss the optical response peculiarities for ultrathin plasmonic films of finite lateral size. Due to their plasma frequency spatial dispersion caused by the spatial confinement of the electron motion, the film dielectric permittivity tensor is spatially dispersive as well and so nonlocal. Such a confinement induced nonlocality can result in peculiar magneto-optical effects. For example, the frequency dependence of the magnetic permeability of the film exhibits a sharp resonance structure shifting to the red as the film aspect ratio increases. The properly tuned ultrathin plasmonic films of finite lateral size can feature the negative refraction effect in the IR frequency range. We discuss how to control these magneto-optical properties and show that they can be tuned by adjusting the film chemical composition, plasmonic material quality, the aspect ratio, and the surroundings of the film.

*Corresponding author. Email: ibondarev@nccu.edu

**INTRODUCTION**

Modern nanofabrication techniques make it possible to produce ultrathin plasmonic films of precisely controlled thickness down to a few monolayers [1–3]. Such films are necessary to create ultrathin multifunctional metasurfaces for advanced applications in optoelectronics, ultrafast information technologies, microscopy, imaging, and sensing, as well as for probing the fundamentals of light-matter interactions at the nanoscale [4–14]. As the film thickness decreases, the strong vertical electron confinement can lead to new confinement related and dimensionality related effects [10,15], which require theory development to understand their role in light-matter interactions and magneto-optical response of thin and ultrathin plasmonic films.

Two of us have recently shown theoretically the plasma frequency of ultrathin films to acquire the spatial dispersion typical of two-dimensional (2D) materials such as graphene, gradually shifting to the red with the film thickness reduction, while being constant for relatively thick films [15]. As a consequence, the complex-valued dynamical dielectric response function shows the gradual red shift of its epsilon-near-zero point with the dissipative loss decreasing at any fixed frequency and gradually going up at the plasma frequency as it shifts to the red with the film thickness reduction. This explains the recent plasma frequency measurements on ultrathin TiN films of controlled variable thickness [3]. The dielectric response *nonlocality* associated with the plasma frequency spatial dispersion we report about is solely a confinement effect. This effect is different from and stronger than the effects commonly known to occur in the framework of hydrodynamical Drude models [5,12,14] due to the pressure term in degenerate electron gas systems [16,17]. Here, we extend our theory to include the analysis of the *magneto-dielectric* response due to the plasma frequency spatial dispersion of ultrathin plasmonic films. We show that the properly tuned ultrathin films of finite lateral size can possess new interesting features such as the resonance magnetic response and the low-frequency negative refraction, which could open up new avenues for potential applications in modern optoelectronics.

**THEORY**

In thin films [see Fig. 1 (a)], the pair Coulomb interaction of charges strengthens with the thickness reduction [18] if the film *background* dielectric constant $\varepsilon$ is much larger than the dielectric constants of the film substrate and superstrate $\varepsilon_{1,2}$. This is because the field produced by the confined charges outside of their confinement region starts playing a perceptible role with its size reduction. When $\varepsilon_{1,2} \ll \varepsilon$ and the in-plane inter-charge distance $\rho$ is greater than the film thickness $d$ as shown in Fig. 1 (a), then the increased 'outside' Coulomb contribution makes the interaction between the charges confined much stronger than that in homogeneous medium with the dielectric constant $\varepsilon$. This leads to the fact that in the finite-thickness metallic films the Fourier transform of the inter-electron Coulomb potential (also known as Keldysh potential [18]) goes as $4\pi e^2/k[kd+(\varepsilon_1+\varepsilon_2)/\varepsilon]$ with $k$ being the absolute value of the 2D (in-plane) electron quasimomentum, and not as $4\pi e^2/k^2$ commonly used, where $k$ is the absolute value of the 3D quasimomentum in bulk materials. In other words, the vertical confinement causes the effective dimensionality reduction and changes the way in which the confined electrons interact with each other. The plasma frequency calculated using this fact takes the following form [15]

$$\omega_p = \omega_p(k) = \frac{\omega_p^{3D}}{\sqrt{1+(\varepsilon_1+\varepsilon_2)/\varepsilon kd}}. \tag{1}$$

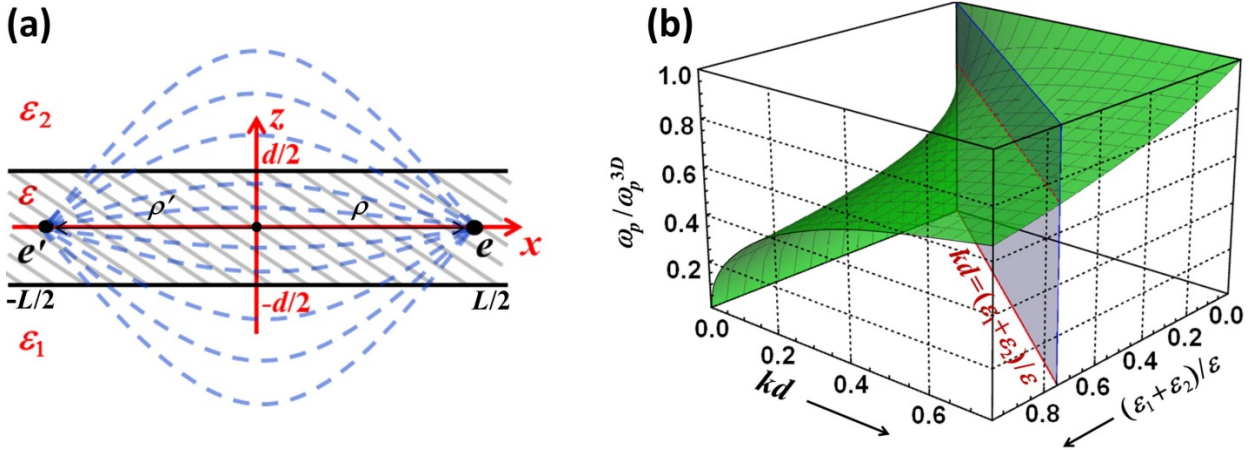

Fig. 1: (a) Schematic of the thin film geometry and sketch of the field line pattern for two confined charges $e$ and $e'$ separated by the in-plane distance $|\boldsymbol{\rho}-\boldsymbol{\rho}'|$ much greater than the film thickness $d$. (b) Thin film plasma frequency normalized by the bulk plasma frequency as given by Eq. (1). The regimes of the relatively thick and ultrathin films are separated by the vertical plane.

Here, $\omega_p^{3D} = \sqrt{4\pi e^2 N_{3D}/\varepsilon m^*}$ is the plasma frequency of the bulk electron gas with the effective mass $m^*$ and volumetric density $N_{3D}$. Equation (1) is presented in Fig. 1 (b). It gives $\omega_p = \omega_p^{3D}$ for relatively thick films with $(\varepsilon_1+\varepsilon_2)/\varepsilon k d \ll 1$, whereas $\omega_p = \omega_p^{2D}(k) = \sqrt{4\pi e^2 N_{2D} k/(\varepsilon_1+\varepsilon_2)m^*}$ in the opposite limit of the ultrathin films with $(\varepsilon_1+\varepsilon_2)/\varepsilon k d \gg 1$ ($N_{2D} = N_{3D}\, d$ is the *surface* electron density), which agrees precisely with the $k^{1/2}$ plasma frequency spatial dispersion of the 2D electron gas in air (see, e.g., Ref. [19]), but does show the explicit dependence on surroundings in the way it occurs for planar monolayer systems such as graphene [20].

### **Spatial dispersion and nonlocality**

The spatially dispersive plasma frequency (1) is contained in the complex-valued dynamical dielectric response function of the electron gas confined in metallic films. This makes the film dielectric permittivity spatially dispersive, i.e. $k$-dependent and so nonlocal. When the spatial dispersion is present, the permittivity is a tensor, not a scalar — even for isotropic layered media shown in Fig. 1 (a) — a distinctive direction is generated by the *in-plane* wave vector **k** [21]. When the medium is not only isotropic but also has a center of symmetry, such a *nonlocal* dielectric response tensor can be written in terms of the **k** components as follows

$$\varepsilon_{\lambda\mu}(\omega,\mathbf{k}) = \varepsilon_t(\omega,k)\left(\delta_{\lambda\mu} - \frac{k_\lambda k_\mu}{k^2}\right) + \varepsilon_l(\omega,k)\frac{k_\lambda k_\mu}{k^2}, \quad \lambda,\mu = x,y \tag{2}$$

where the transverse $\varepsilon_t$ and longitudinal $\varepsilon_l$ response functions depend on $k\,(=|\mathbf{k}|)$ and $\omega$ only [21]. With losses taken into account phenomenologically they can be approximated as

$$\frac{\varepsilon_t(\omega,k)}{\varepsilon} = 1 - \frac{\omega_p^2(k)}{\omega(\omega+i\gamma)}, \qquad \frac{\varepsilon_l(\omega,k)}{\varepsilon} = 1 + \frac{\omega_p^2(k)}{\omega_p^2(k) - \omega(\omega+i\gamma)}, \tag{3}$$

where $\gamma$ is the phenomenological electron inelastic scattering rate and $\omega_p(k)$ is given by Eq. (1). These expressions (the former is commonly referred to as the Drude response function [10]) are normally used to describe the isotropic frequency response of the outer-shell (*s*-band) electrons in metals to the *perpendicularly* and *longitudinally* polarized electromagnetic fields, respectively, with $\varepsilon$ being responsible for the positive background of the ions screened by the remaining inner-shell electrons. In many cases, however, they need to be supplemented with an extra term to take into account the interband electronic transitions. The total response function is then referred to as the Drude-Lorentz function [2,3].

It is interesting to evaluate the degree of nonlocality associated with the plasma frequency spatial dispersion in Eq. (3). This can be done by means of the inverse Fourier transformation from the reciprocal 2D space to the direct coordinate 2D space. For $\varepsilon_t(\omega,k)$, in particular, one has the coordinate-dependent specific permittivity (per unit area) of the form

$$\frac{\varepsilon_t(\boldsymbol{\rho},\boldsymbol{\rho}',\omega)}{\varepsilon} = \delta(|\boldsymbol{\rho}-\boldsymbol{\rho}'|) - \frac{\left(\omega_p^{3D}/2\pi\right)^2}{\omega(\omega+i\gamma)} \int_{k_0}^{k_c} dk\,k \int_0^{2\pi} d\varphi \frac{e^{ik|\boldsymbol{\rho}-\boldsymbol{\rho}'|\cos\varphi}}{1+(\varepsilon_1+\varepsilon_2)/\varepsilon kd}, \tag{4}$$

where $k_0\,(=2\pi/L)$ and $k_c$ represent the plasmon lowest and largest cut-off wave vectors, respectively. For the thick films with $(\varepsilon_1+\varepsilon_2)/\varepsilon k_0 d \ll 1$, this gives the local response of the form

$$\frac{\varepsilon_t(\boldsymbol{\rho},\boldsymbol{\rho}',\omega)}{\varepsilon} = \left[1 - \frac{\left(\omega_p^{3D}\right)^2}{\omega(\omega+i\gamma)}\right]\delta(|\boldsymbol{\rho}-\boldsymbol{\rho}'|). \tag{4a}$$

For the ultrathin films with $(\varepsilon_1+\varepsilon_2)/\varepsilon k_c d \gg 1$, one obtains (in the limit $L\to\infty$)

$$\frac{\varepsilon_t(\boldsymbol{\rho},\boldsymbol{\rho}',\omega)}{\varepsilon} \approx \left[1 - \frac{\varepsilon k_c d\left(\omega_p^{3D}\right)^2}{(\varepsilon_1+\varepsilon_2)\omega(\omega+i\gamma)}\right]\delta(|\boldsymbol{\rho}-\boldsymbol{\rho}'|) + \frac{\varepsilon k_c d\left(\omega_p^{3D}\right)^2}{(\varepsilon_1+\varepsilon_2)\omega(\omega+i\gamma)} \cdot \frac{\cos\left(k_c|\boldsymbol{\rho}-\boldsymbol{\rho}'|+3\pi/4\right)}{\sqrt{2\pi^3 k_c}\,|\boldsymbol{\rho}-\boldsymbol{\rho}'|^{5/2}} \tag{4b}$$

under the extra condition $k_c|\boldsymbol{\rho}-\boldsymbol{\rho}'|\gg 1$ to specify the distances where the Coulomb interaction of the individual electrons is screened and so their motion in the form of the collective plasma oscillations is well defined. One can now see that the ultrathin film permittivity decays with distance as $|\boldsymbol{\rho}-\boldsymbol{\rho}'|^{-5/2}$ to represent the confinement related dielectric response nonlocality that we deal herewith.

### **Dielectric permittivity and magnetic permeability**

At frequencies well below the interband transition frequencies, the description of the linear electromagnetic response of an isotropic (centrosymmetric) medium by means of the spatially dispersive dielectric tensor of Eqs. (2) and (3), which includes responses to both perpendicularly and longitudinally polarized electromagnetic waves, is known to be *equivalent* to the description in terms of the *nondispersive* isotropic dielectric permittivity and magnetic permeability, $\varepsilon(\omega)$ and $\mu(\omega)$ (see Ref. [21]). The equivalency of the two description methods requires

$$\varepsilon(\omega) = \lim_{k\to k_0}\varepsilon_t(\omega,k) = \lim_{k\to k_0}\varepsilon_l(\omega,k) \tag{5}$$

and

$$1 - \frac{1}{\mu(\omega)} = \frac{\omega^2}{c^2}\lim_{k\to k_0}\frac{\varepsilon_t(\omega,k)-\varepsilon_l(\omega,k)}{k^2}. \tag{6}$$

From Eq. (3) one can easily see that Eq. (5) is fulfilled for frequencies $\omega > \omega_p^{3D}\sqrt{2\pi\varepsilon d/(\varepsilon_1+\varepsilon_2)L}$, with the lower boundary controlled by the ratio of the thin film thickness $d$ and lateral size $L$ as shown in Fig. 1 (a), to result in

$$\varepsilon(\omega) = \varepsilon\left\{1 - \frac{\left(\omega_p^{3D}\right)^2}{[1+(\varepsilon_1+\varepsilon_2)L/2\pi\varepsilon d]\omega(\omega+i\gamma)}\right\}. \qquad (7)$$

In the same frequency range, substituting Eq. (3) into Eq. (6), one obtains

$$\mu(\omega) = \left\{1 - \frac{\left(\omega_p^{3D}\right)^2\left(L/\lambda_p^{3D}\right)^2}{[1+(\varepsilon_1+\varepsilon_2)L/2\pi\varepsilon d]^2(\omega-i\gamma)^2}\right\}^{-1}. \qquad (8)$$

For lower frequencies, Eq. (5) is not fulfilled, however, Eqs. (7) and (8) still remain adequate for describing the linear response to the *perpendicularly* polarized electromagnetic waves *alone* [22]. In Eq. (8), the sign of the imaginary part has been manually reversed to reflect the fact that the constitutive relations are differently defined for the electric and magnetic field vectors, $\boldsymbol{E}$ $(=\boldsymbol{D}/\varepsilon)$ and $\boldsymbol{B}$ $(=\mu\boldsymbol{H})$, which represent the fundamental fields (as opposed to the vectors $\boldsymbol{D}$ and $\boldsymbol{H}$ introduced as a matter of convenience [23]) and therefore must have the same time evolution.

Equations (7) and (8) represent the '$\varepsilon(\omega)$ and $\mu(\omega)$' method of the isotropic media description as applied to the ultrathin plasmonic films of finite lateral size. This is the equivalent alternative (under the restrictions abovementioned) to the description based on the spatially dispersive dielectric permittivity tensor defined in Eqs. (2) and (3) where the nonlocal dielectric tensor describes both electric and magnetic field responses [21,22]. For typical experimental values of $L \sim 30\,\mu\text{m}$, $\varepsilon \sim 10$, $\varepsilon_1 \sim \varepsilon_2 \sim 1$, the aspect ratio factor in Eqs. (7) and (8) is estimated to be in the range $(\varepsilon_1+\varepsilon_2)L/2\pi\varepsilon d \sim 10^3 \div 10^2$ for $d \sim 1 \div 10\,\text{nm}$.

## DISCUSSION

Figure 2 presents our model calculations for the real and imaginary parts of $\varepsilon(\omega)$ $(=\varepsilon'+i\varepsilon'')$ and $\mu(\omega)$ $(=\mu'+i\mu'')$ in Eqs. (7) and (8) as functions of the dimensionless frequency $\omega/\omega_p^{3D}$ and the aspect ratio parameter $(\varepsilon_1+\varepsilon_2)L/2\pi\varepsilon d$, with the other parameter values as follows: $\varepsilon = 10$, $\gamma/\omega_p^{3D} = 0.001$, and $L/\lambda_p^{3D} = 100$ (chosen to be close to those of gold [10]). Figures 2 (b) and 2 (d) show the contour plots obtained by cutting the graphs in Figs. 2 (a) and 2 (c) by the parallel vertical planes of constant aspect ratio $(\varepsilon_1+\varepsilon_2)L/2\pi\varepsilon d$, with the thick horizontal blue arrows indicating the aspect ratio increase direction. As the aspect ratio parameter increases, for the dielectric permittivity in Figs. 2 (a) and 2 (b) one can see the red shift of the plasma frequency (zeros of the real parts) accompanied by the dissipative loss reduction (imaginary parts going lower), quite similar to what was earlier reported both experimentally and theoretically [3,15]. The magnetic permeability in Figs. 2 (c) and 2 (d) exhibits a sharp resonance structure shifting to the red as the aspect ratio parameter increases, with the real parts changing sign and the imaginary parts being sharply peaked at zeros of the real parts. These effects are controlled by the aspect ratio parameter $(\varepsilon_1+\varepsilon_2)L/2\pi\varepsilon d$ and, as this parameter includes substrate and superstrate material characteristics, allow the tune-up of the magneto-optical properties of ultrathin plasmonic films not only by varying their chemical composition ($\varepsilon$, $\omega_p^{3D}$) and material quality ($\gamma$) but also by adjusting their $L/d$ ratio (must always be $\gg 1$ though) and by choosing the deposition substrates ($\varepsilon_1$) and coating layers ($\varepsilon_2$) appropriately (with the inequality $\varepsilon \gg \varepsilon_{1,2}$ still to hold).

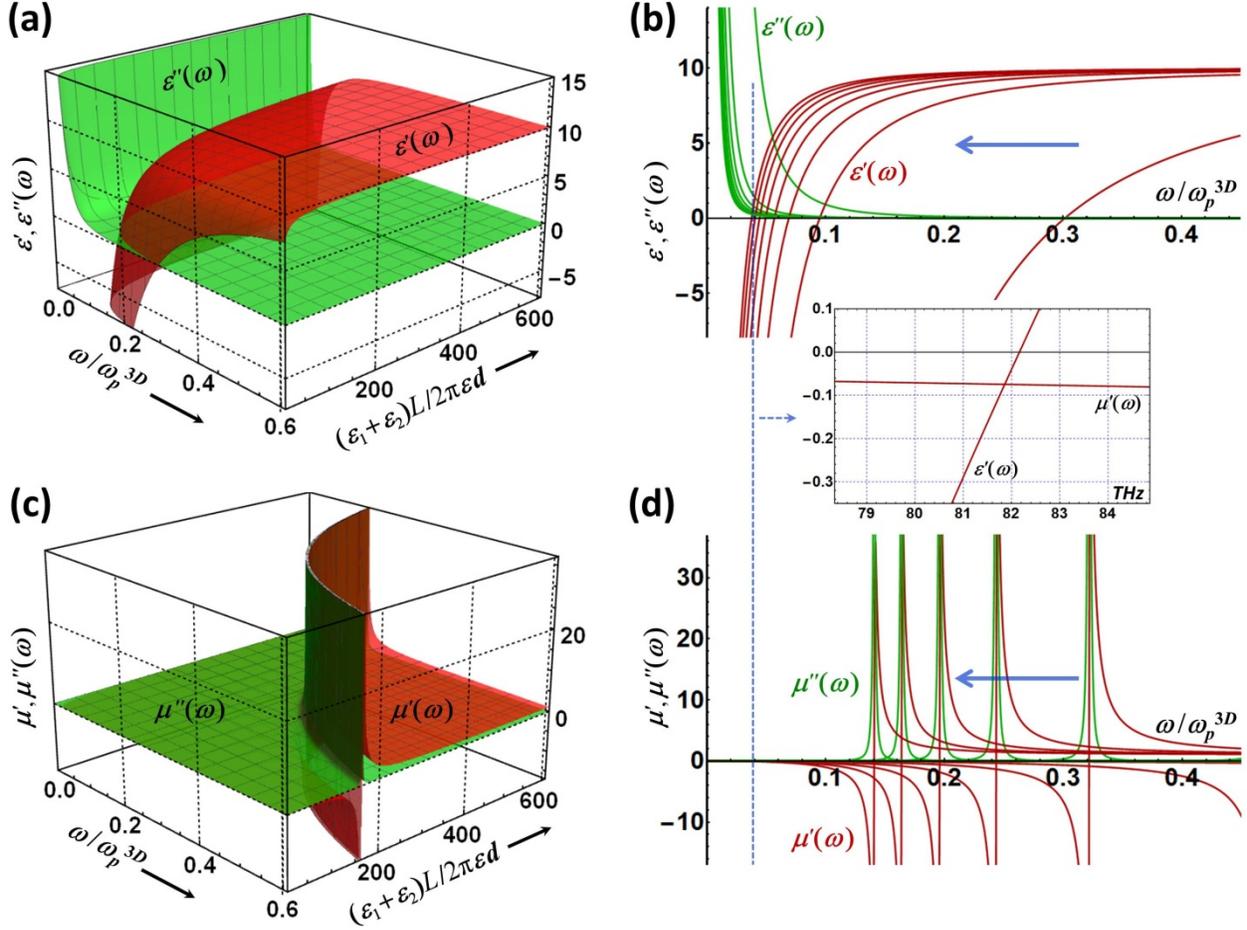

Fig. 2: (a) Real (red) and imaginary (green) parts of the isotropic dielectric permittivity in Eq. (7) as functions of the dimensionless frequency $\omega/\omega_p^{3D}$ and the aspect ratio parameter $(\varepsilon_1+\varepsilon_2)L/2\pi\varepsilon d$ of the film. (b) The contour plot one obtains by cutting the graph in (a) with the parallel vertical planes of constant $(\varepsilon_1+\varepsilon_2)L/2\pi\varepsilon d$. The thick horizontal blue arrow shows the direction of the $(\varepsilon_1+\varepsilon_2)L/2\pi\varepsilon d$ increase. (c,d) Same as in (a,b) for the real and imaginary parts of the isotropic magnetic permeability of Eq. (8). The vertical dashed line going through (b,d) indicates the region where the real part of the dielectric permittivity and the real part of the magnetic permeability are both negative, and the inset shows their actual behavior in this region with frequency expressed in THz. See text for more details.

## **Possibility for low-frequency negative refraction**

When described in terms of the '$\varepsilon(\omega)$ and $\mu(\omega)$' approach, an isotropic linear passive medium is known to possess a negative refractive index provided that the real parts of its dielectric permittivity and magnetic permeability are both negative [24,25]. The vertical dashed line in Figs. 2 (b) and 2 (d) indicates the domain where this could be possible for the ultrathin plasmonic films of finite lateral size. The inset shows the actual behavior of $\varepsilon'(\omega)$ and $\mu'(\omega)$ in this region with the frequency expressed in THz (rescaled from Figs. 2 (b) and 2 (d) with the bulk plasma frequency $\omega_p^{3D}(\text{Au}) \approx 9\,\text{eV} = 2179.19\,\text{THz}$ [10]). One can see that both $\varepsilon'$ and $\mu'$ are negative simultaneously in the IR frequency range. A close inspection of Eqs. (7) and (8) reveals that in order for this to occur, the frequency must satisfy the following inequality

$$\gamma \ll \omega < \omega_p^{3D} \sqrt{\frac{2\pi\varepsilon d}{(\varepsilon_1 + \varepsilon_2)L}}. \tag{9}$$

This condition is exactly opposite to that where the '$\varepsilon(\omega)$ and $\mu(\omega)$' isotropic medium description method of Eqs. (7) and (8) is *equivalent* to the medium description by means of the *spatially* dispersive dielectric tensor of Eqs. (2) and (3), making it possible to describe the linear responses to both perpendicularly and longitudinally polarized electromagnetic waves in terms of $\varepsilon(\omega)$ and $\mu(\omega)$. However, as discussed above, the '$\varepsilon(\omega)$ and $\mu(\omega)$' method still remains adequate at frequencies in Eq. (9) for describing the linear response to the perpendicularly polarized electromagnetic waves alone, that is, for the optical spectroscopy experiment description.

From Eq. (9) one can see that the negative refraction domain of the perpendicularly polarized electromagnetic waves is controlled by $\omega_p^{3D}$ and the inverse aspect ratio parameter of the film on the upper boundary and by the electron inelastic scattering rate on the lower boundary. This domain can be both shrunk and expanded under appropriate material and geometry adjustment. For example, increasing the lateral size of the thin film and/or using a plasmonic material with small $\varepsilon$ and $\omega_p^{3D}$ and relatively large $\gamma$, can shrink the negative refraction domain down to zero. On the other hand, using very clean (small $\gamma$) ultrathin films of finite aspect ratio $L/d$, made of a plasmonic material with large $\varepsilon$ and $\omega_p^{3D}$, placed in low-permittivity surroundings (e.g., air), can expand significantly the negative refraction range for perpendicularly polarized electromagnetic waves. Notably, this possibility only exists for the ultrathin plasmonic films of finite lateral size. In these systems the electron confinement and associated plasma frequency spatial dispersion lead to the dielectric response nonlocality which causes magneto-optical effects such as the negative refraction discussed here and the resonance magnetic response discussed above.

## CONCLUSIONS

In this communication, we discuss some peculiarities of the magneto-dielectric response of ultrathin plasmonic films of finite lateral size. We show that the spatial dispersion and associated nonlocality of the dielectric response, originating from the confinement induced plasma frequency spatial dispersion, can result in new interesting features of the dynamical magnetic response of the film. The frequency dependence of the magnetic permeability of the film features the sharp resonance structure shifting to the red as the film aspect ratio increases. When tuned appropriately, the ultrathin films of finite lateral size can be negatively refractive in the IR frequency range. These features can be tuned by adjusting the film chemical composition ($\varepsilon$, $\omega_p^{3D}$), material quality ($\gamma$) and the $L/d$ ratio as well as by choosing the deposition substrates ($\varepsilon_1$) and coating layers ($\varepsilon_2$) appropriately. Importantly [15], these are all solely confinement related effects different from those commonly known to occur in hydrodynamical Drude models due to the pressure term in the degenerate electron gas with no confinement included [5,12,14]. We believe that our findings open up entirely new avenues for potential applications of ultrathin plasmonic films in modern optoelectronics.

## ACKNOWLEDGEMENTS

I.V.B. is supported by NSF-ECCS-1306871. Work of H.M. is funded by DOE-DE-SC0007117. Work on this project by V.M.S. is supported in part by ONR-N00014-16-1-3003. Discussions with Mikhail Lapine (UT-Sydney, Australia), Alexander Kildishev, Zhaxylyk Kudyshev, and Michael Povolotskyi (all from Purdue University, USA) are gratefully acknowledged.